\documentclass[a4paper,11pt]{article}
\usepackage{jcappub, bm, color} 
\usepackage{amssymb,amsfonts,slashed,amsthm,amsmath,graphicx, soul}
\usepackage[caption=false]{subfig}
\newcommand{\MeV}{\  {\rm MeV} }
\newcommand{\GeV}{\  {\rm GeV} }
\newcommand{\TeV}{\  {\rm TeV} }

\newcommand{\lmk}{\left(}  
\newcommand{\rmk}{\right)}
\newcommand{\lkk}{\left[}  
\newcommand{\rkk}{\right]}

\newcommand{\la}{\left\langle} 
\newcommand{\ra}{\right\rangle}

\newcommand{\bea}{\begin{array}}
\newcommand{\eea}{\end{array}}
\newcommand{\beq}{\begin{eqnarray}}
\newcommand{\eeq}{\end{eqnarray}}

\newcommand{\Mpl}{M_{\rm Pl}}

\newcommand{\abs}[1]{\left\vert {#1} \right\vert}

\newcommand{\Tr}{\text{Tr}}
\newcommand{\Det}{\text{Det}}
\newcommand{\diag}{{\rm diag}}

\newcommand{\nn}{\nonumber\\}

\def\REF#1{Ref.~\cite{#1}}
\def\REFS#1{Refs.~\cite{#1}}
\def\SEC#1{Sec.~\ref{#1}}
\def\APP#1{Appendix~\ref{#1}}
\def\FIG#1{Fig.~\ref{#1}}
\def\TAB#1{Table~\ref{#1}}
\def\EQ#1{Eq.~(\ref{#1})}

\title{
Unification of the Standard Model \\ 
and Dark Matter Sectors 
\\in [SU(5)$\bm{\times}$U(1)]$^{\bm 4}$
}

\author{
Ayuki Kamada,$^{1}$
}
\affiliation{
$^{1}$ Center for Theoretical Physics of the Universe, 
Institute for Basic Science (IBS), 55 Expo-ro, Yuseong-gu, Daejeon 34126, Korea
}

\author{
Masaki Yamada,$^{2}$
}
\affiliation{
$^{2}$ Institute of Cosmology, Department of Physics and Astronomy, 
Tufts University, 574 Boston Avenue, Medford, MA 02155, U.S.A.
}

\author{
Tsutomu T. Yanagida$^{3, 4}$
}
\affiliation{
$^{3}$ T. D.  Lee Institute and School of Physics and Astronomy, Shanghai Jiao Tong University, 800 Dongchuan Rd, Shanghai 200240, China
}
\affiliation{$^{4}$ Kavli IPMU (WPI), UTIAS, 
The University of Tokyo, 5-1-5 Kashiwanoha, Kashiwa, Chiba 277-8583, Japan
}

\abstract{
A simple model of dark matter contains a light Dirac field charged under a hidden U(1) gauge symmetry. When a chiral matter content in a strong dynamics satisfies the t'Hooft anomaly matching condition, a massless baryon is a natural candidate of the light Dirac field. One realization is the same matter content as the standard SU(5)$\times$U(1)$_{(B-L)}$ grand unified theory. We propose a chiral [SU(5)$\times$U(1)]$^4$ gauge theory as a unified model of the SM and DM sectors. The low-energy dynamics, which was recently studied, is governed by the hidden U(1)$_4$ gauge interaction and the third-family U(1)$_{(B-L)_3}$ gauge interaction. This model can realize self-interacting dark matter and alleviate the small-scale crisis of collisionless cold dark matter in the cosmological structure formation. The model can also address the semi-leptonic $B$-decay anomaly reported by the LHCb experiment.
}

\begin{document}

\begin{flushright}
CTPU-PTC-19-11
\end{flushright}

\maketitle
\flushbottom

%%%%%%%%%%%%%%%%%%%%%%%%%%%%%%%%%%%%%%%%%%%%%%%%%%%%%%%%%%%%%%%%
\section{Introduction
\label{sec:introduction}}
%%%%%%%%%%%%%%%%%%%%%%%%%%%%%%%%%%%%%%%%%%%%%%%%%%%%%%%%%%%%%%%%

A simple framework of dark matter (DM) consists of a light Dirac fermion charged under a new U(1) gauge symmetry.
The U(1) symmetry, which is spontaneously broken to some discrete group at low energy, ensures the stability of the light Dirac fermion.
Annihilation of the light Dirac fermions into the gauge bosons determines its thermal relic to be consistent with the observed DM abundance.
Here, one may ask a couple of questions: why is the Dirac fermion light?; and what is the origin of the U(1) gauge boson?
Since the Dirac mass term is allowed by any symmetries, it is mysterious that the mass of Dirac field is as light as, e.g., the electroweak scale. 

We can naturally realize the DM framework based on a chiral SU(5) gauge theory that becomes strong at some intermediate scale.
We introduce two chiral ``preons'' whose representations are $\bm{\bar{5}}$ and $\bm{10}$, which are analogous to the minimal SU(5) grand unified theory (GUT) of the standard model (SM).
Around 1970s, there were a lot of efforts to identify quarks and leptons as composite states of preons~\cite{Pati:1974yy, Terazawa:1976xx, Neeman:1979wp, Harari:1979gi, Shupe:1979fv, Fritzsch:1981zh}.
In this context, t'Hooft showed that the anomaly matching condition must be satisfied when there is a massless composite fermion at low energy~\cite{tHooft:1979rat}.
Although nobody has found a viable theory for composite quarks and leptons, \REF{Dimopoulos:1980hn} found that $\bm{\bar{5}}$ and $\bm{10}$ in a strong SU(5) dynamics results in a massless fermion.
There is a gauge-anomaly-free global U(1) symmetry, which is analogous to the U(1)$_{B-L}$ symmetry in the SM SU(5) GUT.
The [U(1) graviton$^2$] and [U(1)]$^3$ anomalies at high energy and at low energy match.%
\footnote{
Generically, $N$ anti-fundamentals and one antisymmetric tensor in a strong SU($4+N$) dynamics leave $N (N + 1) / 2$ massless composite fermions.
The massless fermions may be identified as right-handed neutrinos~\cite{ArkaniHamed:1998pf}.
See \REF{Gavela:2018paw} for an application of a similar idea to the Peccei-Quinn mechanism.
A model of DM from a strong SU(5) gauge theory was discussed in \REF{Hong:2018hvp}, although vector-like pairs of preons are introduced.
}
By introducing a Dirac partner of the massless fermion, which is analogous to the right-handed neutrino, we gauge the U(1) symmetry.
A dimension 6 operator among those fermions at a high-energy scale results in a Dirac mass term below the dynamical scale. 
The mass scale is of order the electroweak scale when the operator is suppressed by the Planck scale and the dynamical scale is of order $10^{13} \GeV$. 
Therefore, we naturally obtain a light Dirac fermion and a U(1) gauge interaction at low energy from the SU(5)$\times$U(1) gauge theory. 

When the gauge boson is lighter than the U(1)-breaking Higgs, the gauge boson is stable and harmful in cosmology.
A kinetic mixing with some other gauge boson makes it decay to SM particles.
On the other hand, if the kinetic mixing also makes late-time annihilation of DM Dirac fermions result in high-energy electromagnetic particles, it is tightly constrained by high-energy cosmic-ray experiments and observations of cosmic microwave background anisotropies (i.e., indirect detection experiments).
These problems are evaded when the hidden U(1) gauge boson kinetically mixes only with the third-family U(1)$_{(B-L)}$ (i.e., U(1)$_{(B-L)_{3}}$)~\cite{Kamada:2018kmi}.
When the hidden U(1) gauge boson mass is below the tau lepton mass, it dominantly decays to third-family neutrinos.

\begin{figure} %  figure placement: here, top, bottom, or page
   \centering
   \includegraphics[width=4.5in]{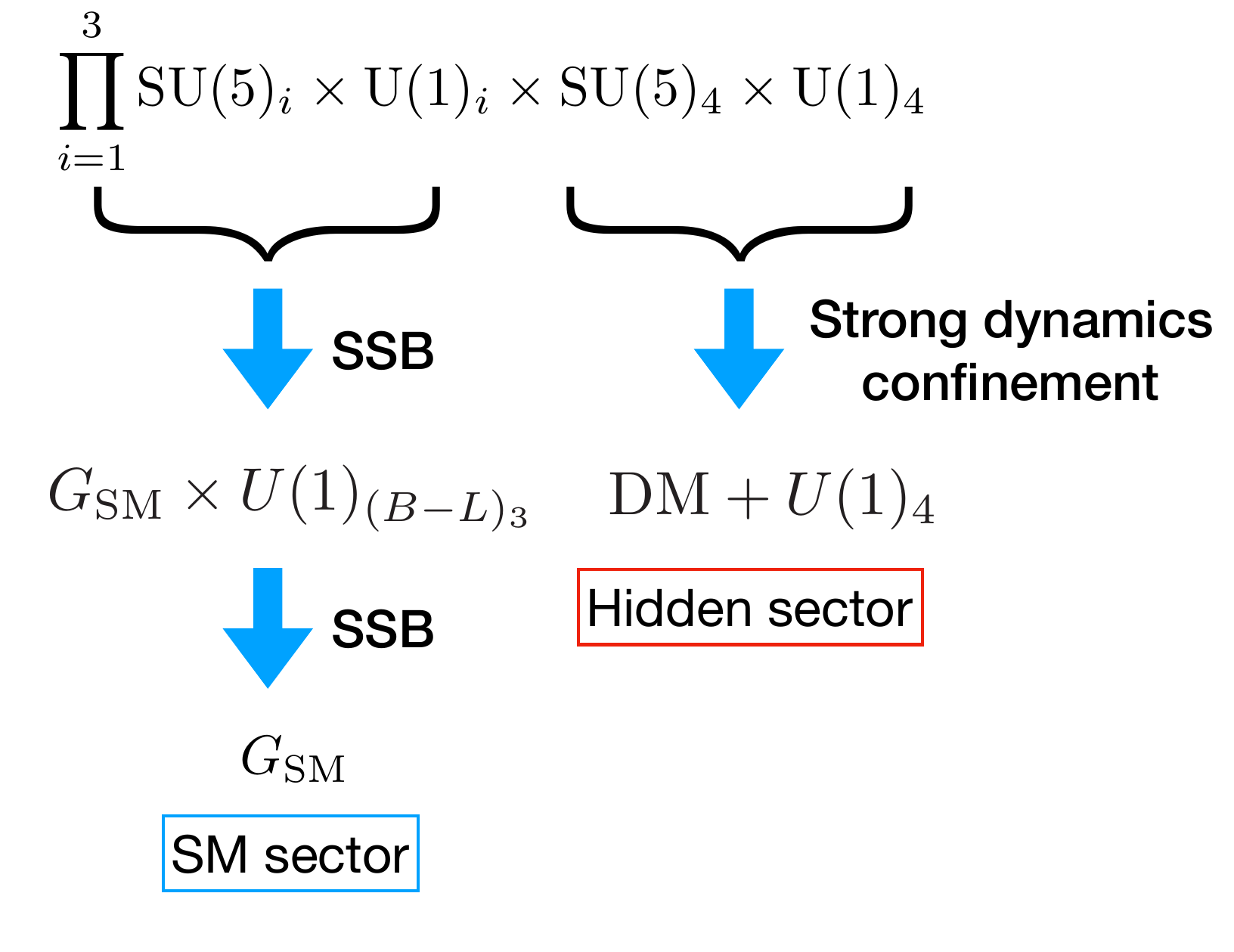} 
   \caption{
   Gauge symmetry breaking pattern in our unified model. 
   The SM gauge group is denoted as $G_{\rm SM} = {\rm SU}(3)_c \times {\rm SU}(2)_L \times {\rm U}(1)_Y$.
   }
   \label{fig:schematic}
\end{figure}

The above observation motivates us to consider a flavored GUT [SU(5)$\times$U(1)]$^{4}$.
Three SM families are charged under the different SU(5)$_i\times$U(1)$_i$ $(i = 1, 2, 3)$.
We identify the chiral SU(5)$_4\times$U(1)$_4$ gauge theory as the DM sector. 
The first three pairs of gauge symmetries are spontaneously broken into the SM gauge symmetries and the last one becomes strong at an intermediate scale to give the SIDM model (see \FIG{fig:schematic}).

Low-energy phenomenology has been studied in \REF{Kamada:2018kmi}. 
The U(1)$_{4}$ gauge symmetry is assumed to be spontaneously broken at $10$-$100 \MeV$.
The U(1)$_{4}$ gauge boson can mediate a strong self-interaction of DM.
Self-interacting dark matter (SIDM) may alleviate tensions of collisionless DM in the small-scale structure formation (see, e.g., \REF{Tulin:2017ara} for a review of SIDM).
For example, SIDM can explain the diversity of galactic rotation curves~\cite{Kamada:2016euw, Creasey:2016jaq, Ren:2018jpt}, while it is not easy for collisionless DM~\cite{Oman:2015xda} (see, e.g., \REF{Bullock:2017xww} for a review of the small-scale crisis).
The self-interaction mediated by the light U(1)$_{4}$ gauge boson is strong at low velocities, but is weakened at high velocities so that it evades constraints from observations of galaxy clusters~\cite{Kaplinghat:2015aga} (see Refs.~\cite{Feng:2009mn, Tulin:2012wi, Dasgupta:2013zpn, Bringmann:2013vra, Ko:2014bka, Cherry:2014xra, Kitahara:2016zyb, Ma:2017ucp, Balducci:2018ryj, Kamada:2018zxi} for SIDM models in a similar direction).
Furthermore, the U(1)$_{(B-L)_{3}}$ gauge boson and flavor mixing, e.g., through the vector-like fermions~\cite{Alonso:2017uky}, introduces a new contribution to semi-leptonic Wilson coefficients.
It can explain the lepton flavor-universality violation in semi-leptonic $B$ meson decays recently reported by the Belle experiment~\cite{Wehle:2016yoi, Abdesselam:2019wac} and LHCb experiment~\cite{Aaij:2014ora, Aaij:2017vbb, Aaij:2019wad} (see also updated analysis in terms of semi-leptonic Wilson coefficients~\cite{Ciuchini:2019usw, Aebischer:2019mlg}). 

This paper is organized as follows. 
In the next section, we briefly explain the simple model of DM.
The model can realize SIDM to address the small-scale crisis of collisionless DM.
The model can also explain the B-decay anomaly.
In \SEC{sec:UV}, we demonstrate that the SM sector as well as the DM sector can be unified in four copies of a SU(5)$\times$U(1) gauge dynamics.
The simple low-energy model of DM can be naturally obtained from the strong dynamics of the high-energy SU(5)$_4 \times$U(1)$_4$ gauge theory.
We specify the bosonic particle content and breaking pattern of SU(5)$_i\times$U(1)$_i$ $(i = 1, 2, 3)$ into the SM to see that the gauge couplings are unified at a high scale.
We conclude in \SEC{sec:conclusions}.

%%%%%%%%%%%%%%%%%%%%%%%%%%%%%%%%%%%%%%%%%%%%%%%%%%%%%%%%%%%%%%%%
\section{Low-energy model
\label{sec:IR}}
%%%%%%%%%%%%%%%%%%%%%%%%%%%%%%%%%%%%%%%%%%%%%%%%%%%%%%%%%%%%%%%%

In this section we briefly review a simple framework of DM introduced in \REF{Kamada:2018kmi}.
The low-energy model is based on U(1)$_{(B-L)_{3}} \times$U(1)$_{4}$ extension of the SM model.
In particular we put stress on that the flavored U(1)$_{(B-L)_{3}}$ symmetry is needed, which leads us to introduce $\prod_{i=1}^3$ [SU(5)$_i\times$U(1)$_i$].
We also briefly describe a mechanism to reproduce the observed flavor mixing.
We assume that the flavor structure is generated by a set of TeV-scale vector-like fermions introduced for each SM fermion (including the right-handed neutrinos).

\subsection{Simple framework of DM}
\label{sec:DM}

In a simple DM framework, we introduce a Dirac field $\eta$ (DM) charged under a hidden gauge symmetry U(1)$_4$.%
\footnote{
In \REF{Kamada:2018kmi}, $\chi$ and U(1)$_h$ denote $\eta$ and U(1)$_4$, respectively.
}
The Lagrangian density in the DM sector is given by 
\beq
 {\cal L}_h = {\rm (kinetic \ terms)} - m_\eta \bar{\eta} \eta  - V_{\Phi_4} (\Phi_4)\,, 
\eeq
where $\Phi_4$ is a Higgs field that spontaneously breaks U(1)$_4$ into some discrete group.
This discrete group stabilizes the Dirac field $\eta$.
Annihilation of $\eta$ into the U(1)$_4$ gauge boson $Z_{4}$ leads to thermal freeze-out of $\eta$ and determines the relic abundance.
$Z_{4}$ also mediates the velocity-dependent self-interaction of $\eta$.
If the U(1)$_4$ is spontaneously broken at $10$-$100 \MeV$, the self-scattering cross section is large enough at low velocities to alleviate the small-scale crisis of collisionless DM.
Meanwhile, its cross section is small enough to evade constraints from galaxy clusters.

If $Z_{4}$ is stable, its relic abundance will overclose the Universe.
To make it decay, we introduce the U(1)$_{(B-L)_3}$ gauge symmetry and a kinetic mixing between the gauge boson $Z_{(B-L)_3}$ and $Z_{4}$:
\beq
 {\cal L}_3 = {\rm (kinetic \ terms)} - \left( \frac12 y_{3} \Phi_3 N_{3} N_{3} + {\rm h.c.} \right) - V_{\Phi_3} 
 (\Phi_3) - \frac12 \epsilon_2 F_{4 \, \mu \nu} F_{(B-L)_3}^{\mu \nu}  \,, 
\eeq
where $F$'s are the field strengths of the corresponding gauge bosons.
$N_3$ is the third-family right-handed neutrino and $\Phi_3$ is a Higgs field that breaks U(1)$_{(B-L)_3}$ spontaneously. 
If the $Z_{4}$ mass is below the tau lepton mass, only the decay channel of $Z_{4}$ into the third-family neutrinos is kinematically allowed.
Indirect detection bounds on DM are systematically weakened because DM annihilation mainly result in neutrinos. 
A kinetic mixing between the hypercharge gauge boson $Z_{Y}$ and $Z_{(B-L)_3}$ is also induced by SM-particle loops:
\beq
 {\cal L}_{Y \text{-} (B-L)_3} = - \frac12 \epsilon_1 F_{Y \, \mu \nu} F_{(B-L)_3}^{\mu \nu} \,.
\eeq
The kinetic mixing leads to observable signals in direct and indirect detection experiments of DM~\cite{Kamada:2018kmi}.

\subsection{Flavor structure and leptogenesis}
\label{sec:FS}

Although it is forbidden by the flavored symmetries U(1)$_{(B-L)_i}$ $(i = 1, 2, 3)$, 
the proper structure of the Yukawa interactions in the SM sector is expected to be induced from their spontaneous symmetry breakings.
We assume that these U(1)$_{(B-L)_i}$ are spontaneously broken by vacuum expectation values (VEVs) $v_{i}$ of Higgs fields $\Phi_i$. 
We consider that $v_{3}$ is at the TeV scale so that the U(1)$_4$ gauge boson can efficiently decay into neutrinos via the kinetic mixing $\epsilon_{2}$. 
We also assume that $v_1$ and $v_2$ are of order $10^9 \GeV$ so that we can realize the seesaw mechanism to explain the small but nonzero masses of SM neutrinos~\cite{Minkowski:1977sc, Yanagida:1979as, GellMann:1980vs, Glashow:1979nm} and the thermal leptogenesis via the decay of the first and second-family right-handed neutrinos $N_1$ and $N_2$~\cite{Fukugita:1986hr} (see, e.g., \REFS{Buchmuller:2002rq, Giudice:2003jh, Buchmuller:2005eh, Davidson:2008bu} for recent reviews). 
We also need another set of vector-like fermions above the $v_1$ and $v_2$ scales to obtain the Yukawa interactions between the first and second families. 

The SM Yukawa matrices can be diagonalized by a unitary rotation for each fermion: $f = U_f f'$ $( f = u_L, d_L, u_R, d_R, \nu_L, l_L, l_R)$.
In the mass eigenstate, the interactions with the $Z_{(B-L)_3}$ gauge boson are then given by 
\beq
 {\cal L} \supset - \sum_f g_{(B-L)_3} Q_f Z^\mu_{(B-L)_3} J_{f,\mu}\,,
 \label{eq:Zinteraction}
 \\
 J_{f,\mu} = \sum_{i,j=1}^3 \bar{f}_i (U_f)^*_{3i} (U_f)_{3j} \gamma_\mu 
 f_j \,,
\eeq
where $Q_f$ is a U(1)$_{(B-L)_3}$ charge of fermion $f$. 
As a result, $Z_{(B-L)_3}$ mediates interactions between different families in the mass eigenstate. 
It is beyond the scope of this paper to study the possible structure of $U_{f}$.
\REF{Kamada:2018kmi} assumes that the CKM and PMNS matrices are attributed to $u_L$ and $\nu_L$, respectively.
It is also assumed that an additional family rotation can appear only between the second and third families.
Under the existence of the additional rotation, the $Z_{4}$ mass should be smaller than the muon mass so that its decay into muons is kinematically prohibited. 
An additional family rotation between the first and third generation should be minuscule to prohibit $Z_{4}$ from significantly decaying into electrons.

The additional family rotation between the second and third families leads to interesting phenomenology in collider physics. 
In particular, we can explain the semi-leptonic $B$ meson decays recently reported by the LHCb experiment~\cite{Aaij:2014ora, Aaij:2017vbb}, consistently with other collider events, like $D^0$-$\bar{D}^0$ mixing and $\tau \to 3 \mu$~\cite{Kamada:2018kmi} (see also \REF{Alonso:2017uky}). 
A benchmark point, where all collider constraints as well as the DM constraints 
are evaded, is 
$\alpha_{(B-L)_3} = 10^{-4}, \alpha_4 = 10^{-3}, m_{Z_{(B-L)_3}} = 70 \GeV, m_{Z_4} = 10 \MeV, m_\eta = 40 \GeV, \epsilon_1 = 10^{-2}$, and $\epsilon_2 = 4 \times 10^{-2}$.

%%%%%%%%%%%%%%%%%%%%%%%%%%%%%%%%%%%%%%%%%%%%%%%%%%%%%%%%%%%%%%%%
\section{High-energy model
\label{sec:UV}}
%%%%%%%%%%%%%%%%%%%%%%%%%%%%%%%%%%%%%%%%%%%%%%%%%%%%%%%%%%%%%%%%

In this section, we explain our unified model for the SM and DM sectors. 
There remain a question unanswered in the low-energy DM model described in the previous section:
why a DM Dirac field is light? and where the almost unbroken U(1)$_4$ symmetry originate?
First, we see that a chiral SU(5)$_4 \times$U(1)$_4$ gauge theory in the DM sector answers these questions.
Then, motivated by the flavored symmetry, we introduce a $\prod_{i=1}^3$ [SU(5)$_i$$\times$U(1)$_i$] gauge theory in the SM sector.
We study the symmetry breaking pattern and gauge coupling unification.
Finally, we end up with a [SU(5)$\times$U(1)]$^4$ gauge theory in a unified model of the SM and DM sectors.

\subsection{DM sector from a SU(5)$_4 \times$U(1)$_4$ GUT}

Let us introduce chiral SU(5)$_4$ dynamics with $\bm{\bar{5}}$ and $\bm{10}$ representations $\psi_4 (-3)$ and $\chi_4 (1)$ (i.e., preons).
The parentheses denote the charges under global U(1) symmetry, which we denote as U(1)$_4$ and is not anomalous with the gauge (except for gravity) symmetries.
We assume that the SU(5)$_4$ gauge interaction becomes strong  and is confined below an intermediate scale $\Lambda$.
At low energy,
there is a chiral baryon composed of three fermions as 
\beq
 \eta \equiv \psi_4 \psi_4 \chi_4 \,. 
\eeq
The U(1)$_4$ charge of this baryon state is $-5$. 
One can check that $\eta$ satisfies the t'Hooft anomaly matching condition of [U(1)$_4$ graviton$^2$] and [U(1)$_4$]$^3$.
Thus $\eta$ can be massless.%
\footnote{
It is also possible that the U(1)$_4$ is dynamically broken by the condensate of $(\psi_4 \psi_4 \chi_4)^2$. 
In this paper, we assume that the condensation scale of operators that break U(1)$_4$ is sufficiently small so that $Z_4$ gauge boson mass can be as light as the electroweak scale. 
}

Since we want to have DM coupled to a light gauge boson, we promote the U(1)$_4$ symmetry to a gauge symmetry.%
\footnote{If the U(1)$_4$ symmetry is just global, quantum gravity effects may induce a Majorana mass to $\eta$.
Although the Majorana mass may be minuscule since it arises from a dimension 9 operator, it is better that we avoid it by gauging the U(1)$_4$ symmetry.}
Once we promote it to the gauge symmetry, we have a gauge anomaly, such as [U(1)$_4$]$^3$. 
However, it is easy to cancel the anomaly by introducing a Dirac partner of the $\eta$, which is denoted by $N_{4}$.
This is nothing but the chiral fermion called the right-handed neutrino in the SM sector.
It is surprising that the matter content in this SU(5)$_4 \times$U(1)$_4$ sector is completely the same as that of the standard SU(5)$\times$U(1)$_{B-L}$ model.
Here we note that the following non-renormalizable operator is allowed by the gauge symmetries: 
\beq
 \frac{c}{\Mpl^2} N_4 \psi_4 \psi_4 \chi_4 + {\rm h.c.} \,,
 \label{eq:Dmass}
\eeq
where we expect a constant $c$ to be of order unity. 
This gives a Dirac mass for $\eta$ and $N_4$ ($\equiv \bar{\eta}$) of order $\Lambda^3 / \Mpl^2$.%
\footnote{If the charge of the U(1)$_4$ breaking Higgs $\Phi_4$ is -10 in analogy to that of U(1)$_{B-L}$, the Majorana mass of order its VEV $v_4$ is introduced to $N_4$.
The introduced mass splitting between DM states changes the DM phenomenology (see, e.g., \REF{Kamada:2018zxi}) from the pure Dirac case studied in \REF{Kamada:2018kmi}, where the U(1)$_4$ charge of $\Phi_4$ is three times that of $\eta$.
}
If $\Lambda = 10^{13} \GeV$, the Dirac mass is around the electroweak scale. 
This Dirac field is identified as DM discussed in \SEC{sec:DM}.

%%%%%%%%%%%%%%%%%%%%%%%%%%%%%%%%%%%%%%%%%%%%%%%%%%%%%%%%%%%%%%%%
\subsection{SM sector from a [SU(5)$\times$U(1)]$^3$ GUT
\label{sec:SMsector}}
%%%%%%%%%%%%%%%%%%%%%%%%%%%%%%%%%%%%%%%%%%%%%%%%%%%%%%%%%%%%%%%%

In a similar way as the minimal SU(5) GUT, we introduce a $\prod_{i=1}^3$ [SU(5)$_i$$\times$U(1)$_i$] theory with $\bm{\bar{5}}$ and $\bm{10}$ representations $\psi_i$ and $\chi_i$ ($i=1,2,3$). 
We also introduce the right-handed neutrinos $N_i$ charged only under U(1)$_i$.
The charge assignment for these fermions is summarized in \TAB{tab:table1}. 
One can check again that these gauge interactions are free from quantum anomaly.

%%%%%%%%%%%%%%%%%%%%%%%%%%%%%%%%%%%%%%%%%%%%%%%%%%%%%%%%%%%%%%%%
{\renewcommand\arraystretch{1.2}
\begin{table}[t]
\begin{center}
\caption{Fermionic field (matter) content (i = 1,2,3,4).
\label{tab:table1}}
\begin{tabular}{p{2cm}p{1.5cm}p{1.5cm}p{1.5cm}p{1.5cm}p{0.75cm}}
\hline
\hline
& $\psi_i$ & $\chi_i$ &$ N_i $ \\
\hline
SU(5)$_i$ & $\bm{\bar{5}}$ & $\bm{10}$  & ${\bm 1}$  \\
U(1)$_i$ & $-3$ & $1$ & $5$  \\
\hline \hline
\end{tabular}\end{center}
\end{table}
}
%%%%%%%%%%%%%%%%%%%%%%%%%%%%%%%%%%%%%%%%%%%%%%%%%%%%%%%%%%%%%%%%

SU(5)$_i$ ($i = 1,2,3$) are assumed to be spontaneously broken into the diagonal subgroup of SU(3)$_c \times$SU(2)$_L \times$U(1)$_Y$.%
\footnote{
This part is similar to a [SO(10)]$^{3}$ GUT~\cite{Babu:2007mb, Alonso:2017uky}. 
}
The U(1)$_i$ symmetry can be identified as U(1)$_{(B-L)_i}$ ($i = 1,2,3$) by taking a linear combination with U(1)$_Y$. 
In the following, we study the breaking pattern and gauge coupling unification.
They are non-trivial because there are many ways to spontaneously break the $\prod_{i=1}^3$SU(5)$_i$ gauge group into the SM gauge group and there may be light fields that affect the renormalization group running of the gauge couplings.
We explain one specific scenario that can realize the gauge coupling unification. 

We note that we do not assume supersymmetry%
\footnote{
If the theory respects supersymmetry at a high energy scale, 
it should be spontaneously broken above $10^{13} \GeV$ 
so that it does not affect the strong dynamics of SU(5)$_4$. 
We note that the supersymmetry may improve the gauge coupling unification without adjoint scalars $\Psi_i$. 
}
nor try to address the doublet-triplet splitting for the Higgs field in this paper. 
We therefore expect that fine-tuning problems can be addressed by some unknown mechanism or may not be a problem at all in quantum field theories. 
In this context, it may be reasonable to assume that there are some light fields much below the GUT scale in addition to the SM Higgs doublet.

We introduce bifundamental scalars $\Omega_{ij}$ ($i \ne j$; $\Omega_{ij} = \Omega_{ji}^{*}$) and adjoint scalars $\Sigma_i$ in addition to Higgs fields $H_i$.
The charge assignment of these scalar fields is summarized in \TAB{tab:table2}.
Since there can be many terms in the scalar potential, we do not give details of the scalar potential.
$\prod_{i=1}^3$[SU(5)$_i$] are spontaneously broken by the VEVs of $\Omega_{ij}$ and $\Sigma_i$ as
\beq
 \label{eq:Omegavev}
&& \la \Omega_{ij} \ra = \diag \lmk a_{ij}, a_{ij}, a_{ij}, b_{ij}, b_{ij} \rmk \,, \\
&& \la \Sigma_i \ra = c_i \cdot \diag \lmk 1/3, 1/3, 1/3, -1/2, -1/2 \rmk \,, 
 \label{eq:Sigmavev}
\eeq
where $a_{ij} \ne b_{ij}$. 
Here we note that $\la \Omega_{ij} \ra$ does not need to be traceless.
In particular, either $a_{ij}$ or $b_{ij}$ can be zero at least at an intermediate scale.
There are several ways to spontaneously break $\prod_{i=1}^3$[SU(5)$_i$] into the SM group $G_{\rm SM}$. 
In the rest of this section, we consider the following sequence of the spontaneous symmetry breakings as an example: 
\beq
 \prod_{i=1}^3[{\rm SU(5)}_i] \to  [{\rm SU}(3)]^2 \times {\rm SU}(2)_L \times {\rm U}(1)_Y
 \to {\rm SU}(3)_c \times {\rm SU}(2)_L \times {\rm U}(1)_Y \,.
 \label{eq:SSB}
\eeq
We denote the energy scales of the first and second spontaneous symmetry breakings as $E_{\rm GUT}$ and $E_{\rm SU(3)}$ $(\ll E_{\rm GUT})$, respectively. 
This can be realized when $a_{12}$, $b_{12}$, and $b_{23}$ are about $E_{\rm GUT}$, while $a_{23}$ is about $E_{\rm SU(3)}$.%
\footnote{
We note that all of $a_{ij}$ and $b_{ij}$ do not need to have a nonzero VEV to break [SU(5)]$^3$ into the SM group. Here we assume that $a_{31}$ ($b_{31}$) is smaller than $a_{12}$ and $a_{23}$ ($b_{12}$ and $b_{23}$) or zero. 
}
A further discussion on the scalar potential is given in \APP{sec:scalarV}.
We denote $[{\rm SU}(3)]^2$ at the intermediate scale as SU(3)$_d \times$SU(3)$_3$, where SU(3)$_d$ is a diagonal subgroup of SU(5)$_1\times$SU(5)$_2$ and SU(3)$_3$ is a subgroup of SU(5)$_3$.
In this example, the colored bifundamental $\Omega_{c, 31}$ from $\Omega_{31}$ spontaneously breaks SU(3)$_d \times$SU(3)$_3$ at $E_{\rm SU(3)}$.
The Yukawa interactions between the first and second families can be introduced at $E_{\rm GUT}$.%
\footnote{
Here we implicitly assume that U(1)$_i$ ($i=1,2$) are spontaneously broken at the same scale for simplicity. 
If the U(1)$_i$ breaking scale is lighter, the set of vector-like multiplets of SU(5)$_i$ should also be lighter to produce the Yukawa interactions between the first and second families. 
Since they form complete multiplets of SU(5)$_i$, they do not spoil the unification of the gauge couplings. 
}

We note that the renormalizable potential (e.g., $\Tr [\Omega_{12} \Omega_{23} \Omega_{31}]$) accidentally respects global relative-phase rotations among $\Omega_{ij}$.
However, they can be explicitly broken by dimension 5 operators like $\Det [\Omega_{ij}]$.
The associated pseudo-NG bosons are therefore much heavier than the electroweak scale and are decoupled from the low-energy physics. 
We assume that one of the mass eigenstate among the Higgs doublets $H_{L, i}$ is at the electroweak scale so that we can spontaneously break SU(2)$_L \times$U(1)$_Y$ down to U(1)$_{\rm em}$.

As commented in \SEC{sec:FS}, there should be a non-trivial field content at the TeV scale to reproduce a correct Yukawa structure. 
Following \REF{Alonso:2017uky}, we consider that there is a set of vector-like fermions introduced for each SM fermion at the TeV scale. 
We also need another set of vector-like fermions above the U(1)$_i$ ($i=1,2$) breaking scale to obtain the Yukawa interactions between the first and second families. 
Assuming a universality of particle contents in the ultraviolet physics for SU(5)$_i$  ($i=1,2,3$), we introduce three sets of vector-like multiplets that are ${\bm 5}+\bm{\bar{5}}$ and $\bm{10}+\bm{\overline{10}}$ representations under SU(5)$_i$, respectively. 
Only one set of vector multiplets should be at the TeV scale. 
We identify it as that charged under SU(5)$_3$. 

\begin{figure} %  figure placement: here, top, bottom, or page
   \centering
   \includegraphics[width=4.5in]{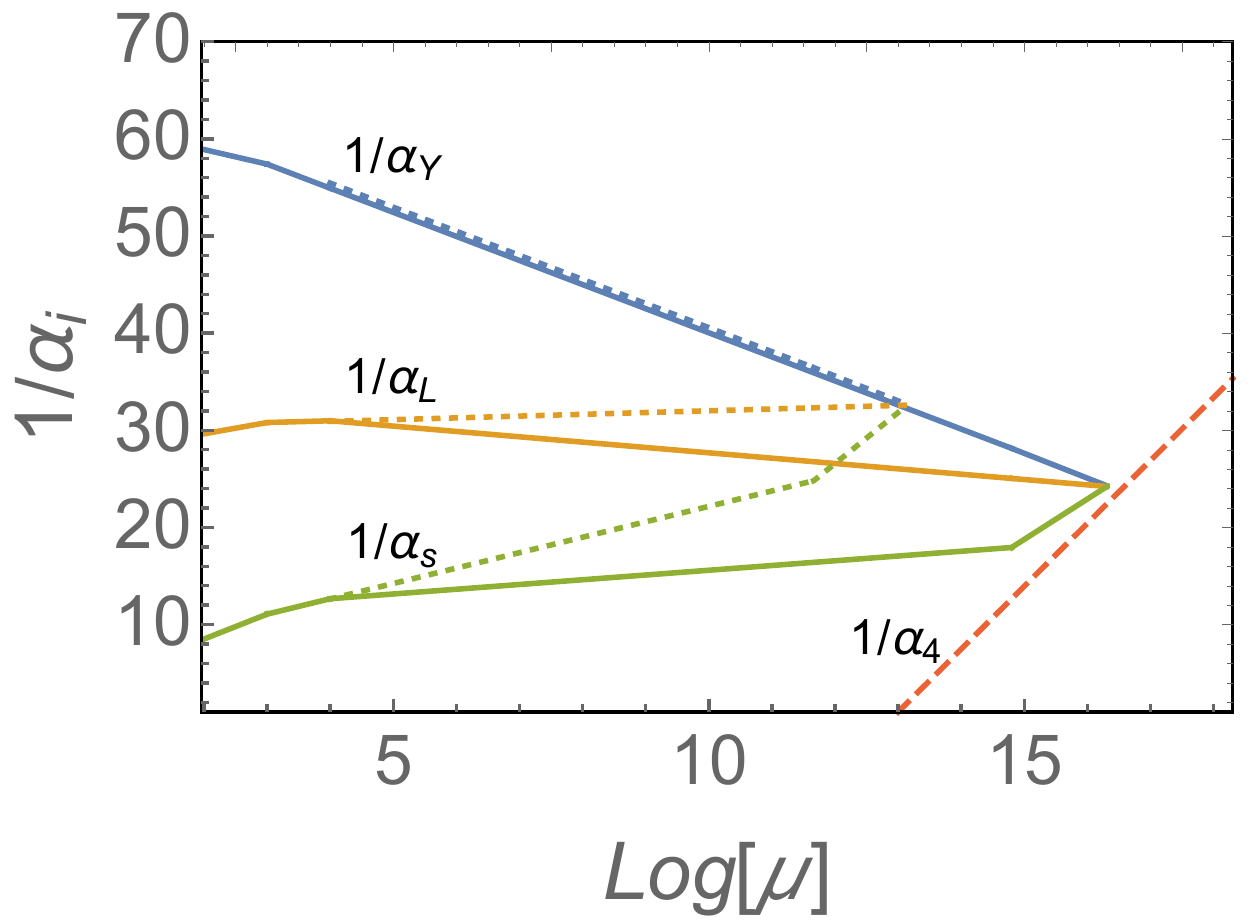} 
   \caption{
   Renormalization group runnings of the gauge couplings with (solid lines) and without (dotted lines) adjoint fields $\Psi_{L, i}$ and $\Psi_{c, i}$ at $10 \TeV$. 
   The renormalization scale $\mu$ is in units of GeV.
   The dashed line represents the running of $\alpha_4^{-1}$. 
   At the energy scale between $E_{\rm SU(3)}$ and $E_{\rm GUT}$, we plot $\alpha_{s}^{-1} = \alpha_{{\rm SU}(3)_d}^{-1} + \alpha_{{\rm SU}(3)_3}^{-1}$ as an effective coupling that should match the SU(3)$_c$ at $E_{\rm SU(3)}$. 
   }
   \label{fig:unification}
\end{figure}

Dotted lines in \FIG{fig:unification} depict the runnings of the gauge couplings.
We summarize details on the runnings of the gauge couplings in \APP{sec:running}.
At $E_{\rm GUT} \simeq 2 \times 10^{13} \GeV$, the gauge group is spontaneously broken to SU(3)$_d \times$SU(3)$_3 \times$SU(2)$_L \times $U(1)$_Y$. 
Then SU(3)$_d \times$SU(3)$_3$ gauge group is spontaneously broken to SU(3)$_c$ at $E_{\rm SU(3)} \simeq 5 \times 10^{11} \GeV$. 
However, our model predicts proton decay that is mediated by the heavy gauge fields. 
The mass scale of the gauge fields is of order $E_{\rm GUT} \simeq 2 \times 10^{13} \GeV$, which is so low that the model is excluded by the constraint on the proton decay rate.

%%%%%%%%%%%%%%%%%%%%%%%%%%%%%%%%%%%%%%%%%%%%%%%%%%%%%%%%%%%%%%%%
{\renewcommand\arraystretch{1.2}
\begin{table}[t]
\begin{center}
\caption{Scalar field content ($i,j = 1,2,3$ and $i \ne j$).
\label{tab:table2}}
\begin{tabular}{p{2cm}p{1.8cm}p{1.8cm}p{1.8cm}p{1.8cm}p{0.75cm}}
\hline
\hline
& $\Omega_{ij}= \Omega_{ji}^{*}$ & $H_i$ &$ \Psi_i $ &$ \Sigma_i $ &$ \Phi_i $ \\
\hline
SU(5)$_i$ & $\bm{\bar{5}}$ & $\bm{\bar{5}}$ & $\bm{24}$ & $\bm{24}$ & ${\bm 1}$   \\
SU(5)$_j$ & $\bm{5}$ & $\bm{1}$  & ${\bm 1}$ & ${\bm 1}$ & ${\bm 1}$ \\
U(1)$_i$ & $0$ & $2$ & $0$ & $0$ & $-10$  \\
\hline \hline
\end{tabular}\end{center}
\end{table}
}
%%%%%%%%%%%%%%%%%%%%%%%%%%%%%%%%%%%%%%%%%%%%%%%%%%%%%%%%%%%%%%%%

To evade the constraint, we introduce three adjoint scalars for SU(3)$_c$ and SU(2)$_L$, $\Psi_{c, i}$ and $\Psi_{L, i}$ $(i = 1, 2, 3)$, at $10\TeV$.
They make the GUT scale larger than that without these fields~\cite{Cox:2016epl}. 
We expect that these fields originate from three scalar fields with the adjoint scalars $\Psi_{i}$. 
The resulting runnings of the gauge couplings are shown as the solid lines in \FIG{fig:unification}. 
The GUT scale is now given by $E_{\rm GUT}  \simeq 2 \times 10^{16} \GeV$ and the SU(3)$_d \times$SU(3)$_3$ breaking scale is given by $E_{\rm SU(3)} \simeq 6 \times 10^{14} \GeV$. 
The GUT scale is high enough to evade the constraint on the proton decay rate. 

The decay of the additional adjoint fields, which we denote as $\Psi_{L, i}$ and $\Psi_{c, i}$, does not cause a cosmological problem.
We can write operators like $\Psi_{L, i} H_{L, i}^\dagger H_{L, i}$ and $\Psi_{c, i} H_{c, i}^\dagger H_{c, i}$, where $H_{c, i}$ are colored Higgs fields. 
The SU(2)$_L$ adjoint fields $\Psi_{L, i}$ decay fast into the SM Higgs field by the former operator.
The SU(3)$_c$ adjoint fields $\Psi_{c, i}$ decay into quarks via one-loop effect, which is suppressed by the mass of colored Higgs field. Since the mass of the colored Higgs fields can be as light as $10^{12} \GeV$%
\footnote{
Light colored Higgs fields changes the running of the gauge couplings only slightly.
}
to evade the constraint on the proton decay rate, the SU(3)$_c$ adjoint field can decay long before the big bang nucleosynthesis epoch. 

Before closing the section, we briefly comment on another simple possibility to realize the gauge coupling unification in our model. 
If the $\prod_{i=1}^3$SU(5)$_i$ gauge group is spontaneously broken into the SM gauge group at a single energy scale, we can use a scenario of gauge coupling unification for the standard SU(5) GUT. 
One simple example was proposed in \REF{Murayama:1991ah}, where we introduced leptoquarks and a $\bm{\bar{5}}$ Higgs field at the TeV scale to realize the gauge coupling unification. 
Although the GUT scale of this model is too low, we can make it high enough by introducing adjoint fields for SU(2)$_L$ and SU(3)$_c$~\cite{Cox:2016epl}.

 %%%%%%%%%%%%%%%%%%%%%%%%%%%%%%%%%%%%%%%%%%%%%%%%%%%%%%%%%%%%%%%%
\subsection{[SU(5)$\times$U(1)]$^4$ unified model
\label{sec:model}}
%%%%%%%%%%%%%%%%%%%%%%%%%%%%%%%%%%%%%%%%%%%%%%%%%%%%%%%%%%%%%%%%
 
So far we discuss a $\prod_{i=1}^3$ [SU(5)$_i$$\times$U(1)$_i$] $(i = 1, 2, 3)$ theory in the SM sector and a SU(5)$_4 \times$U(1)$_4$ theory in the DM sector.
It is surprising that the fermion content in each [SU(5)$_i$$\times$U(1)$_i$] $(i = 1, 2, 3, 4)$ is identical to each other.
In \FIG{fig:unification}, we can see that all SU(5)$_i$ gauge couplings are of the same order, which may indicate that all SU(5)$_i$ gauge groups are unified at a higher scale.
We end up with a unified chiral [SU(5)$\times$U(1)]$^4$ theory with the identical gauge couplings.
We would like to emphasize that the DM mass is predicted to be of order $100 \GeV$ in this unified theory.
The unified gauge coupling is usually in the range of $1/20$-$1/25$ and the GUT scale is about $10^{16} \GeV$ in most reasonable GUT models. 
The dynamical scale of the fourth SU(5)$_4$ gauge theory is then of order $10^{13} \GeV$ because of the running of the gauge coupling from the GUT scale. 
As a result, the mass of DM is then naturally at the TeV scale from the dimension 6 operator. 
The DM mass being predicted is an outstanding feature.
Here we would remember the reason why weakly interacting massive particles (WIMPs) have been attractive.
This is because its mass is predicted at the TeV scale so that TeV-scale new physics solves the naturalness problem.
Thus, we could say that our DM candidate is as compelling (and miraculous) as traditional WIMPs.

%%%%%%%%%%%%%%%%%%%%%%%%%%%%%%%%%%%%%%%%%%%%%%%%%%%%%%%%%%%%%%%%
\section{Conclusion
\label{sec:conclusions}}
%%%%%%%%%%%%%%%%%%%%%%%%%%%%%%%%%%%%%%%%%%%%%%%%%%%%%%%%%%%%%%%%

We have proposed a chiral $\prod_{i=1}^4$ [SU(5)$_i \times$U(1)$_i$]$^4$ gauge theory as a unified model of the SM and DM sectors.
The chiral matter content in each SU(5)$\times$U(1) sector is the same as that of the standard SU(5) GUT. 
The first three sets of SU(5) gauge groups are spontaneously broken into a diagonal subgroup, which is identified as the SM gauge group. 
The first two U(1) gauge symmetries are also spontaneously broken at a high-energy scale, while the third one is spontaneously broken at the TeV scale. 
We have provided an example of the particle content that realizes the gauge coupling unification around $10^{16} \GeV$. 

The fourth SU(5) gauge interaction becomes strong at $\Lambda \sim 10^{13} \GeV$, when its gauge coupling strength is identical to that of the first three sets of SU(5) (i.e., the visible GUT) at the GUT scale.
A chiral baryon state satisfies the t'Hooft anomaly matching condition below the confinement scale.
The baryon and the U(1)$_4$ charged fermion form a Dirac field whose mass, arising from a dimension 6 operator, is of order $\Lambda^3 / \Mpl^2 = 1 \TeV$.
This Dirac field is identified as a dominant component of DM.
In other words, we have found that the model naturally predicts the DM mass of order the electroweak scale from the running of the unified gauge coupling.
Since the unified gauge coupling as well as the GUT scale do not change much in viable GUT scenarios, this is a universal prediction of our model.

The low-energy phenomenology is studied in \REF{Kamada:2018kmi}.
The U(1)$_4$ gauge symmetry is spontaneously broken at low energy into some discrete group, which stabilizes the DM Dirac field.
Thermal freeze-out through annihilations into the U(1)$_4$ gauge bosons reproduces the correct DM abundance.
The U(1)$_4$ gauge boson also mediates the DM self-interaction.
In particular, when the U(1)$_4$ gauge symmetry is spontaneously broken at $10$-$100 \MeV$, the strength and velocity dependence of the self-interaction can be consistent with that indicated by the small-scale crisis of collisionless DM in the cosmological small-scale structure formation. 

We need to introduce the kinetic mixing between U(1)$_4$ and U(1)$_{(B-L)_3}$ gauge boson so that the unwanted U(1)$_4$ gauge boson dominantly decays into the third-family neutrino. 
Through the loop-induced kinetic mixing between the U(1)$_{(B-L)_3}$ and U(1)$_Y$ gauge bosons, still, it can decay into electrons although it is subominant.
It leads to a detectable signal in the indirect detection experiment of DM.
The kinetic mixings also predict observable signals in the direct detection experiment of DM. 
In addition, the lepton flavor-universality violation of semi-leptonic $B$ meson decays reported by the LHCb experiment can be explained by the U(1)$_{(B-L)_3}$ gauge boson with an additional second-third family mixing.

%---------------SECTION------------------%
%
\section*{Acknowledgments}
%
%---------------SECTION------------------%

A.~K. was supported by Institute for Basic Science under the project code, IBS-R018-D1.
A.~K. thanks to Takumi Kuwahara for useful discussions.
T.~T.~Y. was supported in part by the China Grant for Talent Scientific Start-Up Project and the JSPS Grant-in-Aid for Scientific Research No.~16H02176, and No.~17H02878 and by World Premier International Research Center Initiative (WPI Initiative), MEXT, Japan.
T.~T.~Y. thanks to Hamamatsu Photonics.

\appendix
\section{More on the scalar potential}
\label{sec:scalarV}

In this appendix, we give a further discussion on the scalar potential.

\subsection{GUT breaking at hierarchical scales}
\label{sec:scalarVhierarchy}
We consider the symmetry breaking pattern of \EQ{eq:SSB} with hierarchical scales in the main text.
[SU(5)]$^3$ is broken into [SU(3)]$^2\times$SU(2)$_L \times$U(1)$_Y$ at $E_{\rm GUT}$ 
and then into the SM gauge group at $E_{\rm SU(3)}$ ($\ll E_{\rm GUT}$). 

We consider the potential of $\Omega_{ij}$ such as 
\beq
 V_\Omega 
 &=& 
 \sum_{\substack{ (i, j) = \\ (1, 2), (2, 3), (1, 3) }} \lmk 
 M_{ij}^2 \Tr \lkk \Omega_{ij} \Omega_{ji} \rkk 
 + \lambda_{ij} \Tr \lkk \Sigma_i \Omega_{ij} \Sigma_j \Omega_{ji} \rkk 
 + \lambda'_{ij} \Tr \lkk (\Omega_{ij} \Omega_{ji})^2 \rkk 
 \rmk \,, 
\label{eq:potentialOmega2}
\eeq
where we omit the other renormalizable terms for the sale of notational simplicity. 
First, the adjoint scalars $\Sigma_i$ develop VEVs of \EQ{eq:Sigmavev} as in the standard SU(5) GUT model.
These break [SU(5)]$^3$ down to [SU(3)$\times$SU(2)$\times$U(1)]$^3$. 
The of $\Omega_{ij}$ can be rewritten as the effective potential for $a_{ij}$ and $b_{ij}$ defined in \EQ{eq:Omegavev}: 
\beq
 &&V(a_{ij}) = 
  \sum_{\substack{ (i, j) = \\ (1, 2), (2, 3), (1, 3) }} 3\lkk 
 M_{ij,a}^2 a_{ij}^2 + \lambda'_{ij} a_{ij}^4 \rkk \,,
\\
 &&V(b_{ij}) = 
  \sum_{\substack{ (i, j) = \\ (1, 2), (2, 3), (1, 3) }} 2\lkk 
 M_{ij,b}^2 b_{ij}^2 + \lambda'_{ij} b_{ij}^4 \rkk \,, 
\eeq
where 
\beq
 &&M_{ij,a}^2 = M_{ij}^2 + \frac{\lambda_{ij} c_i c_j}{9} \,, 
 \\
 &&M_{ij,b}^2 = M_{ij}^2 + \frac{\lambda_{ij} c_i c_j}{4} \,. 
\eeq

It is possible that $M_{12,a}^2 < 0, M_{12,b}^2 <0, M_{23,a}^2 >0, M_{23,b}^2 < 0, M_{31,a}^2 < 0, M_{31,b}^2 > 0$. 
Then $a_{23}$ and $b_{31}$ are zero while $a_{12}$, $b_{12}$, $b_{23}$, and $a_{31}$ have nonzero values. 
We assume that $a_{12}$, $b_{12}$, and $b_{23}$ are about the GUT scale, $E_{\rm GUT}$, 
while $a_{31}$ is about $E_{\rm SU(3)}$, which is orders of magnitude smaller than $E_{\rm GUT}$. 
This can be realized by taking $\abs{M_{31,a}^2 } \ll \abs{M_{12,a}^2}, \abs{M_{12,b}^2}, \abs{M_{23,b}^2}$ by tuning a parameter. 
In this example, [SU(5)]$^3$ (or [SU(3)$\times$SU(2)$\times$U(1)]$^3$) is broken into SU(3)$\times$SU(3)$\times$SU(2)$_L \times$U(1)$_Y$ at $E_{\rm GUT}$ and then into the SM gauge group at $E_{\rm SU(3)}$. 

\subsection{GUT breaking at a single scale}
\label{sec:scalarVsingle}
We can also consider that [SU(5)]$^3$ is broken to the SM gauge group at a single scale, while we do not consider it in the main text.
In this case, we do not need to introduce the adjoint scalars $\Sigma_i$.
The general potential of $\Omega_{ij}$ is given by 
\beq
 V_\Omega 
 &=& 
 \sum_{\substack{ (i, j) = \\ (1, 2), (2, 3), (1, 3) }} M_{ij}^2 \Tr \lkk \abs{ \Omega_{ij}} ^2 \rkk 
 + \left( A  \, \Tr \lkk \Omega_{12} \Omega_{23} \Omega_{31} \rkk  +  {\rm h.c.} \right) 
 \nn
 &&+ \sum_{\substack{ (i, j, k, l) = \\ (1, 2, 1, 2), (2, 3, 2, 3), (3, 1, 3, 1), \\ (1, 2, 1, 3), (2, 3, 2, 1), (3, 1, 3, 2) }} 
  \lambda_{ijkl} \Tr \Omega_{ij} \Omega_{jk} \Omega_{kl} \Omega_{li}  
  + \sum_{\substack{ (i, j), (k, l) = \\ (1, 2), (2, 3), (1, 3) }} 
  \lambda_{ij,kl} \Tr \abs{\Omega_{ij}}^2 \Tr \abs{\Omega_{kl}}^2 \,,
  \nn
\label{potentialOmega}
\eeq
where $M_{ij}$ are components of a mass matrix, $A > 0$ is a parameter with mass dimension one, 
and $\lambda_{ijkl}$ are quartic coupling constants. 

Although the explicit values of these VEVs are quite complicated, 
we write them for the case of an universal $\lambda_{ijkl}$ $(\equiv \lambda)$ 
and $\lambda_{ij,kl} = 0$ 
in units of $M_{ij}^2 = -c^2$ as an illustration.
We consider
\beq
 \la \Omega_{ij} \ra = {\rm diag} \lmk a^{1}_{ij}, a^{2}_{ij}, a^{3}_{ij}, a^{4}_{ij}, a^{5}_{ij} \rmk \,,
\eeq
and find
\beq
 && a^{k}_{12} = a^{k}_{23} = a^{k}_{31} \,, \quad a^{k}_{12} = - a^{k}_{23} = - a^{k}_{31} \,, \quad - a^{k}_{12} = a^{k}_{23} = - a^{k}_{31} \,, ~ {\rm or} \quad - a^{k}_{12} = - a^{k}_{23} = a^{k}_{31} \notag \\
 && = - \frac{c}{8 \lambda} \lmk A + \sqrt{A^2 + 16 \lambda} \rmk \,,
\eeq
for $k = 1, 2, 3, 4, 5$.
There are degenerate vacua with SU(5), SU(4)$\times$U(1), or SU(3)$\times$SU(2)$\times$U(1), which are diagonal subgroups of SU(5)$_1 \times $SU(5)$_2 \times$SU(5)$_3$.

\section{Runnings of gauge couplings}
\label{sec:running}
In this section, we summarize the one-loop beta function coefficients $b_i$ $(i = 1,2,3,4)$ used in \FIG{fig:unification}:
\beq
 &&\frac{d g_i}{d \ln \mu} = \frac{b_i}{16 \pi^2} g_i^3 \,.
  \label{eq:RG}
\eeq

The beta function coefficient of the gauge coupling for SU(5)$_4$ is given by 
\beq
 &&b_4 = -\frac{11}{3} \cdot 5 + \frac{2}{3} \cdot \lmk \frac12 + \frac32 \rmk = - \frac{51}{3} \,. 
\eeq
We assume that the dynamical scale of SU(5)$_4$ is of order $10^{13} \GeV$ so that the Dirac mass \EQ{eq:Dmass} is around the electroweak scale. 
The resulting renormalization group running for $\alpha_4$ is shown in \FIG{fig:unification} as the dotted line.

At the energy scale between $E_{\rm SU(3)}$ and $E_{\rm GUT}$, the beta function coefficients of the gauge couplings are as follows: 
\beq
 && \frac{5}{3} b_{{\rm U}(1)_Y} = 
  \frac23 \cdot (3+2) \cdot \lmk 6 \cdot \frac{1}{36} + 3 \cdot \frac49 + 3 \cdot \frac19 + 2 \cdot \frac14 + 1 \rmk
 + \frac13 \cdot \lmk 2 \cdot \frac14 \rmk
 = \frac{203}{18} \,,
 \\
 &&b_{{\rm SU}(2)_L} = - \frac{11}{3} \cdot 2 + \frac{2}{3} \cdot (3+2) \cdot \lmk 3 \cdot \frac12 + \frac12 \rmk
 + \frac{1}{3} \cdot \lmk \frac12  \rmk
 = -\frac{1}{2} \,, 
 \\
 &&b_{{\rm SU}(3)_d} 
 = - \frac{11}{3} \cdot 3+ \frac{2}{3} \cdot 2 \cdot \lmk 2 \cdot \frac12 + \frac12 + \frac12 \rmk + \frac{1}{3} \cdot \lmk 3 \cdot \frac12 \rmk 
 = - \frac{47}{6} \,, 
 \\
 &&b_{{\rm SU}(3)_3}
 = - \frac{11}{3} \cdot 3 + \frac{2}{3} \cdot (1+2) \cdot \lmk 2 \cdot \frac12 + \frac12 + \frac12 \rmk + \frac{1}{3} \cdot \lmk 3 \cdot \frac12  \rmk 
 = - \frac{13}{2} \,. 
\eeq
Here, we take into account the bifundamental scalar $\Omega_{c, 31}$ that spontaneously breaks SU(3)$_d \times$SU(3)$_3$ to SU(3)$_c$ at $\mu = E_{\rm SU(3)}$.

Below $E_{\rm SU(3)}$ but above the TeV scale, the beta function coefficients are given by 
\beq 
 &&\frac{5}{3} b_{{\rm U}(1)_Y} = 
   \frac23 \cdot (3+2) \cdot \lmk 6 \cdot \frac{1}{36} + 3 \cdot \frac49 + 3 \cdot \frac19 + 2 \cdot \frac14 + 1 \rmk
 + \frac13 \cdot \lmk 2 \cdot \frac14 \rmk  = \frac{203}{18}\,,
 \\
 &&b_{{\rm SU}(2)_L} 
 = - \frac{11}{3} \cdot 2 + \frac23 \cdot (3+2) \cdot \lmk 3 \cdot \frac12 + \frac12 \rmk
 + \frac13 \cdot \lmk \frac12 \rmk 
 = - \frac{1}{2} \,,
 \\
 &&b_{{\rm SU}(3)_c} 
 = - \frac{11}{3} \cdot 3 + \frac23 \cdot (3+2) \cdot \lmk 2 \cdot \frac12 + \frac12 + \frac12 \rmk 
 = - \frac{13}{3}\,,
\eeq
where the running is modified by the presence of the vector-like multiplets when compared to the SM. 

The junction conditions of the gauge couplings are given by 
\beq
 &&\alpha_Y^{-1} = \alpha_L^{-1} = \sum_{i=1}^3 \alpha_i^{-1} \,,
 \\
 &&\alpha_{{\rm SU}(3)_d}^{-1} = \sum_{i=1}^2 \alpha_i^{-1}\,,
 \\
 &&\alpha_{{\rm SU}(3)_3}^{-1} = \alpha_3^{-1}\,, 
\eeq
at $\mu = E_{\rm GUT}$ and 
\beq
 \alpha_{s}^{-1} = \alpha_{{\rm SU}(3)_d}^{-1} + \alpha_{{\rm SU}(3)_3}^{-1}\,, 
\eeq
at $\mu = E_{{\rm SU}(3)}$. 
At the energy scale between $E_{\rm SU(3)}$ and $E_{\rm GUT}$, we plot the running of $\alpha_Y^{-1}$, $\alpha_L^{-1}$, and $\alpha_{{\rm SU}(3)_d}^{-1} + \alpha_{{\rm SU}(3)_3}^{-1}$. 
Because of the junction conditions, all of them must be unified at $\mu = E_{\rm GUT}$ and the last combination of the gauge couplings is continuously connected to $\alpha_s^{-1}$ at $\mu = E_{\rm SU(3)}$. 

We introduce three adjoint scalars for SU(3)$_c$ and SU(2)$_L$, $\Psi_{c, i}$ and $\Psi_{L, i}$ $(i = 1, 2, 3)$, at the $10\TeV$ scale to make $E_{\rm GUT}$ higher.
They change the beta function coefficients as follows.
At the energy scale between $E_{\rm SU(3)}$ and $E_{\rm GUT}$,
\beq 
  &&b'_{{\rm SU}(2)_L} = b_{{\rm SU}(2)_L}
  + \frac13 \cdot 3 \cdot \lmk 2 \rmk \,,
 \\
 &&b'_{{\rm SU}(3)_d} = b_{{\rm SU}(3)_d}
 + \frac13 \cdot 2 \cdot \lmk 3 \rmk \,,
 \\
 &&b'_{{\rm SU}(3)_3} = b_{{\rm SU}(3)_3}
 + \frac13 \cdot \lmk 3 \rmk \,.
\eeq
Below $E_{\rm SU(3)}$ but above the TeV scale, 
\beq 
  &&b'_{{\rm SU}(2)_L} = b_{{\rm SU}(2)_L}
  + \frac13 \cdot 3 \cdot \lmk 2 \rmk \,,
 \\
 &&b'_{{\rm SU}(3)_c} = b_{{\rm SU}(3)_c}
 + \frac13 \cdot 3 \cdot \lmk 3 \rmk \,.
\eeq

\bibliography{reference}

\providecommand{\href}[2]{#2}\begingroup\raggedright\begin{thebibliography}{10}

\bibitem{Pati:1974yy}
J.~C. Pati and A.~Salam, \emph{{Lepton Number as the Fourth Color}},
  \href{https://doi.org/10.1103/PhysRevD.10.275,
  10.1103/PhysRevD.11.703.2}{\emph{Phys. Rev.} {\bfseries D10} (1974) 275}.

\bibitem{Terazawa:1976xx}
H.~Terazawa, K.~Akama and Y.~Chikashige, \emph{{Unified Model of the
  Nambu-Jona-Lasinio Type for All Elementary Particle Forces}},
  \href{https://doi.org/10.1103/PhysRevD.15.480}{\emph{Phys. Rev.} {\bfseries
  D15} (1977) 480}.

\bibitem{Neeman:1979wp}
Y.~Ne'eman, \emph{{Irreducible Gauge Theory of a Consolidated Weinberg-Salam
  Model}}, \href{https://doi.org/10.1016/0370-2693(79)90521-5}{\emph{Phys.
  Lett.} {\bfseries B81} (1979) 190}.

\bibitem{Harari:1979gi}
H.~Harari, \emph{{A Schematic Model of Quarks and Leptons}},
  \href{https://doi.org/10.1016/0370-2693(79)90626-9}{\emph{Phys. Lett.}
  {\bfseries 86B} (1979) 83}.

\bibitem{Shupe:1979fv}
M.~A. Shupe, \emph{{A Composite Model of Leptons and Quarks}},
  \href{https://doi.org/10.1016/0370-2693(79)90627-0}{\emph{Phys. Lett.}
  {\bfseries 86B} (1979) 87}.

\bibitem{Fritzsch:1981zh}
H.~Fritzsch and G.~Mandelbaum, \emph{{Weak Interactions as Manifestations of
  the Substructure of Leptons and Quarks}},
  \href{https://doi.org/10.1016/0370-2693(81)90626-2}{\emph{Phys. Lett.}
  {\bfseries 102B} (1981) 319}.

\bibitem{tHooft:1979rat}
G.~'t~Hooft, \emph{{Naturalness, chiral symmetry, and spontaneous chiral
  symmetry breaking}},
  \href{https://doi.org/10.1007/978-1-4684-7571-5_9}{\emph{NATO Sci. Ser. B}
  {\bfseries 59} (1980) 135}.

\bibitem{Dimopoulos:1980hn}
S.~Dimopoulos, S.~Raby and L.~Susskind, \emph{{Light Composite Fermions}},
  \href{https://doi.org/10.1016/0550-3213(80)90215-1}{\emph{Nucl. Phys.}
  {\bfseries B173} (1980) 208}.

\bibitem{ArkaniHamed:1998pf}
N.~Arkani-Hamed and Y.~Grossman, \emph{{Light active and sterile neutrinos from
  compositeness}},
  \href{https://doi.org/10.1016/S0370-2693(99)00672-3}{\emph{Phys. Lett.}
  {\bfseries B459} (1999) 179}
  [\href{https://arxiv.org/abs/hep-ph/9806223}{{\ttfamily hep-ph/9806223}}].

\bibitem{Gavela:2018paw}
M.~B. Gavela, M.~Ibe, P.~Quilez and T.~T. Yanagida, \emph{{Automatic
  Peccei-Quinn symmetry}},  \href{https://arxiv.org/abs/1812.08174}{{\ttfamily
  1812.08174}}.

\bibitem{Hong:2018hvp}
D.~K. Hong, \emph{{A model of light dark matter and dark radiation}},
  \href{https://arxiv.org/abs/1808.10149}{{\ttfamily 1808.10149}}.

\bibitem{Kamada:2018kmi}
A.~Kamada, M.~Yamada and T.~T. Yanagida, \emph{{Self-interacting dark matter
  with a vector mediator: kinetic mixing with U(1)$_{(B-L)_3}$ gauge boson}},
  \href{https://arxiv.org/abs/1811.02567}{{\ttfamily 1811.02567}}.

\bibitem{Tulin:2017ara}
S.~Tulin and H.-B. Yu, \emph{{Dark Matter Self-interactions and Small Scale
  Structure}}, \href{https://doi.org/10.1016/j.physrep.2017.11.004}{\emph{Phys.
  Rept.} {\bfseries 730} (2018) 1}
  [\href{https://arxiv.org/abs/1705.02358}{{\ttfamily 1705.02358}}].

\bibitem{Kamada:2016euw}
A.~Kamada, M.~Kaplinghat, A.~B. Pace and H.-B. Yu, \emph{{How the
  Self-Interacting Dark Matter Model Explains the Diverse Galactic Rotation
  Curves}}, \href{https://doi.org/10.1103/PhysRevLett.119.111102}{\emph{Phys.
  Rev. Lett.} {\bfseries 119} (2017) 111102}
  [\href{https://arxiv.org/abs/1611.02716}{{\ttfamily 1611.02716}}].

\bibitem{Creasey:2016jaq}
P.~Creasey, O.~Sameie, L.~V. Sales, H.-B. Yu, M.~Vogelsberger and J.~Zavala,
  \emph{{Spreading out and staying sharp -- creating diverse rotation curves
  via baryonic and self-interaction effects}},
  \href{https://doi.org/10.1093/mnras/stx522}{\emph{Mon. Not. Roy. Astron.
  Soc.} {\bfseries 468} (2017) 2283}
  [\href{https://arxiv.org/abs/1612.03903}{{\ttfamily 1612.03903}}].

\bibitem{Ren:2018jpt}
T.~Ren, A.~Kwa, M.~Kaplinghat and H.-B. Yu, \emph{{Reconciling the Diversity
  and Uniformity of Galactic Rotation Curves with Self-Interacting Dark
  Matter}},  \href{https://arxiv.org/abs/1808.05695}{{\ttfamily 1808.05695}}.

\bibitem{Oman:2015xda}
K.~A. Oman et~al., \emph{{The unexpected diversity of dwarf galaxy rotation
  curves}}, \href{https://doi.org/10.1093/mnras/stv1504}{\emph{Mon. Not. Roy.
  Astron. Soc.} {\bfseries 452} (2015) 3650}
  [\href{https://arxiv.org/abs/1504.01437}{{\ttfamily 1504.01437}}].

\bibitem{Bullock:2017xww}
J.~S. Bullock and M.~Boylan-Kolchin, \emph{{Small-Scale Challenges to the
  $\Lambda$CDM Paradigm}},
  \href{https://doi.org/10.1146/annurev-astro-091916-055313}{\emph{Ann. Rev.
  Astron. Astrophys.} {\bfseries 55} (2017) 343}
  [\href{https://arxiv.org/abs/1707.04256}{{\ttfamily 1707.04256}}].

\bibitem{Kaplinghat:2015aga}
M.~Kaplinghat, S.~Tulin and H.-B. Yu, \emph{{Dark Matter Halos as Particle
  Colliders: Unified Solution to Small-Scale Structure Puzzles from Dwarfs to
  Clusters}}, \href{https://doi.org/10.1103/PhysRevLett.116.041302}{\emph{Phys.
  Rev. Lett.} {\bfseries 116} (2016) 041302}
  [\href{https://arxiv.org/abs/1508.03339}{{\ttfamily 1508.03339}}].

\bibitem{Feng:2009mn}
J.~L. Feng, M.~Kaplinghat, H.~Tu and H.-B. Yu, \emph{{Hidden Charged Dark
  Matter}}, \href{https://doi.org/10.1088/1475-7516/2009/07/004}{\emph{JCAP}
  {\bfseries 0907} (2009) 004}
  [\href{https://arxiv.org/abs/0905.3039}{{\ttfamily 0905.3039}}].

\bibitem{Tulin:2012wi}
S.~Tulin, H.-B. Yu and K.~M. Zurek, \emph{{Resonant Dark Forces and Small Scale
  Structure}},
  \href{https://doi.org/10.1103/PhysRevLett.110.111301}{\emph{Phys. Rev. Lett.}
  {\bfseries 110} (2013) 111301}
  [\href{https://arxiv.org/abs/1210.0900}{{\ttfamily 1210.0900}}].

\bibitem{Dasgupta:2013zpn}
B.~Dasgupta and J.~Kopp, \emph{{Cosmologically Safe eV-Scale Sterile Neutrinos
  and Improved Dark Matter Structure}},
  \href{https://doi.org/10.1103/PhysRevLett.112.031803}{\emph{Phys. Rev. Lett.}
  {\bfseries 112} (2014) 031803}
  [\href{https://arxiv.org/abs/1310.6337}{{\ttfamily 1310.6337}}].

\bibitem{Bringmann:2013vra}
T.~Bringmann, J.~Hasenkamp and J.~Kersten, \emph{{Tight bonds between sterile
  neutrinos and dark matter}},
  \href{https://doi.org/10.1088/1475-7516/2014/07/042}{\emph{JCAP} {\bfseries
  1407} (2014) 042} [\href{https://arxiv.org/abs/1312.4947}{{\ttfamily
  1312.4947}}].

\bibitem{Ko:2014bka}
P.~Ko and Y.~Tang, \emph{{$\nu \Lambda$MDM: A model for sterile neutrino and
  dark matter reconciles cosmological and neutrino oscillation data after
  BICEP2}}, \href{https://doi.org/10.1016/j.physletb.2014.10.035}{\emph{Phys.
  Lett.} {\bfseries B739} (2014) 62}
  [\href{https://arxiv.org/abs/1404.0236}{{\ttfamily 1404.0236}}].

\bibitem{Cherry:2014xra}
J.~F. Cherry, A.~Friedland and I.~M. Shoemaker, \emph{{Neutrino Portal Dark
  Matter: From Dwarf Galaxies to IceCube}},
  \href{https://arxiv.org/abs/1411.1071}{{\ttfamily 1411.1071}}.

\bibitem{Kitahara:2016zyb}
T.~Kitahara and Y.~Yamamoto, \emph{{Protophobic Light Vector Boson as a
  Mediator to the Dark Sector}},
  \href{https://doi.org/10.1103/PhysRevD.95.015008}{\emph{Phys. Rev.}
  {\bfseries D95} (2017) 015008}
  [\href{https://arxiv.org/abs/1609.01605}{{\ttfamily 1609.01605}}].

\bibitem{Ma:2017ucp}
E.~Ma, \emph{{Inception of Self-Interacting Dark Matter with Dark Charge
  Conjugation Symmetry}},
  \href{https://doi.org/10.1016/j.physletb.2017.06.067}{\emph{Phys. Lett.}
  {\bfseries B772} (2017) 442}
  [\href{https://arxiv.org/abs/1704.04666}{{\ttfamily 1704.04666}}].

\bibitem{Balducci:2018ryj}
O.~Balducci, S.~Hofmann and A.~Kassiteridis, \emph{{Flavor structures in the
  Dark Standard Model TeV-Paradigm}},
  \href{https://arxiv.org/abs/1810.07198}{{\ttfamily 1810.07198}}.

\bibitem{Kamada:2018zxi}
A.~Kamada, K.~Kaneta, K.~Yanagi and H.-B. Yu, \emph{{Self-interacting dark
  matter and muon $g-2$ in a gauged U$(1)_{L_{\mu} - L_{\tau}}$ model}},
  \href{https://doi.org/10.1007/JHEP06(2018)117}{\emph{JHEP} {\bfseries 06}
  (2018) 117} [\href{https://arxiv.org/abs/1805.00651}{{\ttfamily
  1805.00651}}].

\bibitem{Alonso:2017uky}
R.~Alonso, P.~Cox, C.~Han and T.~T. Yanagida, \emph{{Flavoured $B-L$ local
  symmetry and anomalous rare $B$ decays}},
  \href{https://doi.org/10.1016/j.physletb.2017.10.027}{\emph{Phys. Lett.}
  {\bfseries B774} (2017) 643}
  [\href{https://arxiv.org/abs/1705.03858}{{\ttfamily 1705.03858}}].

\bibitem{Wehle:2016yoi}
{\scshape Belle} collaboration, S.~Wehle et~al., \emph{{Lepton-Flavor-Dependent
  Angular Analysis of $B\to K^\ast \ell^+\ell^-$}},
  \href{https://doi.org/10.1103/PhysRevLett.118.111801}{\emph{Phys. Rev. Lett.}
  {\bfseries 118} (2017) 111801}
  [\href{https://arxiv.org/abs/1612.05014}{{\ttfamily 1612.05014}}].

\bibitem{Abdesselam:2019wac}
{\scshape Belle} collaboration, A.~Abdesselam et~al., \emph{{Test of lepton
  flavor universality in ${B\to K^\ast\ell^+\ell^-}$ decays at Belle}},
  \href{https://arxiv.org/abs/1904.02440}{{\ttfamily 1904.02440}}.

\bibitem{Aaij:2014ora}
{\scshape LHCb} collaboration, R.~Aaij et~al., \emph{{Test of lepton
  universality using $B^{+}\rightarrow K^{+}\ell^{+}\ell^{-}$ decays}},
  \href{https://doi.org/10.1103/PhysRevLett.113.151601}{\emph{Phys. Rev. Lett.}
  {\bfseries 113} (2014) 151601}
  [\href{https://arxiv.org/abs/1406.6482}{{\ttfamily 1406.6482}}].

\bibitem{Aaij:2017vbb}
{\scshape LHCb} collaboration, R.~Aaij et~al., \emph{{Test of lepton
  universality with $B^{0} \rightarrow K^{*0}\ell^{+}\ell^{-}$ decays}},
  \href{https://doi.org/10.1007/JHEP08(2017)055}{\emph{JHEP} {\bfseries 08}
  (2017) 055} [\href{https://arxiv.org/abs/1705.05802}{{\ttfamily
  1705.05802}}].

\bibitem{Aaij:2019wad}
{\scshape LHCb} collaboration, R.~Aaij et~al., \emph{{Search for
  lepton-universality violation in $B^+\to K^+\ell^+\ell^-$ decays}},
  \href{https://arxiv.org/abs/1903.09252}{{\ttfamily 1903.09252}}.

\bibitem{Ciuchini:2019usw}
M.~Ciuchini, A.~M. Coutinho, M.~Fedele, E.~Franco, A.~Paul, L.~Silvestrini
  et~al., \emph{{New Physics in $b \to s \ell^+ \ell^-$ confronts new data on
  Lepton Universality}},  \href{https://arxiv.org/abs/1903.09632}{{\ttfamily
  1903.09632}}.

\bibitem{Aebischer:2019mlg}
J.~Aebischer, W.~Altmannshofer, D.~Guadagnoli, M.~Reboud, P.~Stangl and D.~M.
  Straub, \emph{{B-decay discrepancies after Moriond 2019}},
  \href{https://arxiv.org/abs/1903.10434}{{\ttfamily 1903.10434}}.

\bibitem{Minkowski:1977sc}
P.~Minkowski, \emph{{$\mu \to e\gamma$ at a Rate of One Out of $10^{9}$ Muon
  Decays?}}, \href{https://doi.org/10.1016/0370-2693(77)90435-X}{\emph{Phys.
  Lett.} {\bfseries 67B} (1977) 421}.

\bibitem{Yanagida:1979as}
T.~Yanagida, \emph{{HORIZONTAL SYMMETRY AND MASSES OF NEUTRINOS}}, {\emph{Conf.
  Proc.} {\bfseries C7902131} (1979) 95}.

\bibitem{GellMann:1980vs}
M.~Gell-Mann, P.~Ramond and R.~Slansky, \emph{{Complex Spinors and Unified
  Theories}}, {\emph{Conf. Proc.} {\bfseries C790927} (1979) 315}
  [\href{https://arxiv.org/abs/1306.4669}{{\ttfamily 1306.4669}}].

\bibitem{Glashow:1979nm}
S.~L. Glashow, \emph{{The Future of Elementary Particle Physics}},
  \href{https://doi.org/10.1007/978-1-4684-7197-7_15}{\emph{NATO Sci. Ser. B}
  {\bfseries 61} (1980) 687}.

\bibitem{Fukugita:1986hr}
M.~Fukugita and T.~Yanagida, \emph{{Baryogenesis Without Grand Unification}},
  \href{https://doi.org/10.1016/0370-2693(86)91126-3}{\emph{Phys. Lett.}
  {\bfseries B174} (1986) 45}.

\bibitem{Buchmuller:2002rq}
W.~Buchmuller, P.~Di~Bari and M.~Plumacher, \emph{{Cosmic microwave background,
  matter - antimatter asymmetry and neutrino masses}},
  \href{https://doi.org/10.1016/S0550-3213(02)00737-X,
  10.1016/j.nuclphysb.2007.11.030}{\emph{Nucl. Phys.} {\bfseries B643} (2002)
  367} [\href{https://arxiv.org/abs/hep-ph/0205349}{{\ttfamily
  hep-ph/0205349}}].

\bibitem{Giudice:2003jh}
G.~F. Giudice, A.~Notari, M.~Raidal, A.~Riotto and A.~Strumia, \emph{{Towards a
  complete theory of thermal leptogenesis in the SM and MSSM}},
  \href{https://doi.org/10.1016/j.nuclphysb.2004.02.019}{\emph{Nucl. Phys.}
  {\bfseries B685} (2004) 89}
  [\href{https://arxiv.org/abs/hep-ph/0310123}{{\ttfamily hep-ph/0310123}}].

\bibitem{Buchmuller:2005eh}
W.~Buchmuller, R.~D. Peccei and T.~Yanagida, \emph{{Leptogenesis as the origin
  of matter}},
  \href{https://doi.org/10.1146/annurev.nucl.55.090704.151558}{\emph{Ann. Rev.
  Nucl. Part. Sci.} {\bfseries 55} (2005) 311}
  [\href{https://arxiv.org/abs/hep-ph/0502169}{{\ttfamily hep-ph/0502169}}].

\bibitem{Davidson:2008bu}
S.~Davidson, E.~Nardi and Y.~Nir, \emph{{Leptogenesis}},
  \href{https://doi.org/10.1016/j.physrep.2008.06.002}{\emph{Phys. Rept.}
  {\bfseries 466} (2008) 105}
  [\href{https://arxiv.org/abs/0802.2962}{{\ttfamily 0802.2962}}].

\bibitem{Babu:2007mb}
K.~S. Babu, S.~M. Barr and I.~Gogoladze, \emph{{Family Unification with
  SO(10)}}, \href{https://doi.org/10.1016/j.physletb.2008.01.057}{\emph{Phys.
  Lett.} {\bfseries B661} (2008) 124}
  [\href{https://arxiv.org/abs/0709.3491}{{\ttfamily 0709.3491}}].

\bibitem{Cox:2016epl}
P.~Cox, A.~Kusenko, O.~Sumensari and T.~T. Yanagida, \emph{{SU(5) Unification
  with TeV-scale Leptoquarks}},
  \href{https://doi.org/10.1007/JHEP03(2017)035}{\emph{JHEP} {\bfseries 03}
  (2017) 035} [\href{https://arxiv.org/abs/1612.03923}{{\ttfamily
  1612.03923}}].

\bibitem{Murayama:1991ah}
H.~Murayama and T.~Yanagida, \emph{{A viable SU(5) GUT with light leptoquark
  bosons}}, \href{https://doi.org/10.1142/S0217732392000070}{\emph{Mod. Phys.
  Lett.} {\bfseries A7} (1992) 147}.

\end{thebibliography}\endgroup

\end{document}